\documentclass{elsart}
 \usepackage{amssymb}
 \usepackage{epsfig}
    \newcommand{\ba}{\begin{eqnarray}}
    \newcommand{\ea}{\end{eqnarray}}
    \newcommand{\be}{\begin{equation}}
    \newcommand{\ee}{\end{equation}}

    \newcommand {\bk} {{\mathbf k}}

    \newcommand {\calM} {{\mathcal M}}
    
    \newcommand{\AmS}{{\protect\the\textfont2
  A\kern-.1667em\lower.5ex\hbox{M}\kern-.125emS}}

\begin{document}
\runauthor{Junhua and Chuan}
\begin{frontmatter}

\title{Quenched Charmed Meson Spectra using Tadpole Improved Quark Action on
Anisotropic Lattices\thanksref{fund}}
\author[Beida]{Liuming Liu},
\author[Beida]{Shiquan Su},
\author[Beida]{Xin Li}
\author[Beida]{and Chuan Liu}
\address[Beida]{School of Physics\\
          Peking University\\
          Beijing, 100871, P.~R.~China}
\thanks[fund]{This work is supported by the Key Project of National Natural
 Science Foundation of China (NSFC) under grant
 No. 10421003, No. 10235040 and supported by the
 Trans-century fund from Chinese Ministry of Education.}

 \begin{abstract}
 Charmed meson charmonium spectra are studied with improved quark actions on
 anisotropic lattices. We measured the pseudo-scalar and vector
 meson dispersion relations for $4$ lowest lattice momentum modes
 with quark mass values ranging from the strange quark to charm quark
 with $3$ different values of gauge coupling $\beta$ and $4$
 different values of bare speed of light $\nu$. With the
 bare speed of light parameter $\nu$ tuned in a mass-dependent way,
 we study the mass spectra of $D$, $D_s$, $\eta_c$,
 $D^{\ast}$, $D_s^{\ast}$ and $J/\psi$ mesons.
 The results extrapolated to the continuum limit are
 compared with the experiment and qualitative agreement is found.
 \end{abstract}
 \end{frontmatter}


\section{Introduction}

 In recent years, anisotropic lattices and improved
 lattice actions have been used extensively in lattice
 QCD calculations. In this letter, we present our preliminary
 numerical study on the spectrum of charmed mesons and charmoniums.
 The gauge action employed in this work is the tadpole improved
 gluonic action on asymmetric lattices as in Ref.~\cite{colin99}:
 Using such a gauge action, glueball and light hadron spectrum has been studied
 within the quenched approximation \cite{colin97,colin99,%
 chuan01:gluea,chuan01:glueb,chuan01:canton1,chuan01:canton2,chuan01:india}.

 It has been suggested that relativistic heavy quarks
 can also be treated with the help of anisotropic lattices
 (the Fermi lab approach), with improvements \cite{kronfeld97:aniso,%
 klassen98:wilson_quark,klassen99:aniso_wilson,kronfeld01:aniso_Oa,%
 harada02:aniso}.
 Using various versions of the quark actions, charmed meson
 spectrum, charmonium spectrum, charmed baryons have
 been studied on the
 lattice~\cite{mackenzie98:BsDs,chen01:aniso,lewis01:aniso,%
 CPPACS02:aniso,juettner03:Ds,nemeto03:aniso_baryon}.
 However, in order to take full advantage of the
 improved quark action on anisotropic lattices, some parameters
 in the action have to be tuned, either perturbatively or
 non-perturbatively, in order to gain as much improvement as possible.
 Some numerical studies of these parameters have already appeared in
 the literature~\cite{chuan01:tune,harada02:aniso,CPPACS03:aniso_unquench,%
 okamoto03:aniso,peardon04:aniso}. In particular, it turns out
 that the bare speed of light parameter $\nu$ has to be tuned
 non-perturbatively in a quark {\em mass-dependent}
 way as is shown in Ref.~\cite{chuan05:S2C_tune}.
 With the parameter $\nu$ tuned as such, we can study quenched
 meson spectrum with quark mass values ranging from the strange
 to the charm, using the same quark action.

 The anisotropic quark actions used in our study is:
 $S_f=\sum_{xy}\bar{\psi}_x\calM_{xy}\psi_y$
 with the fermion matrix given by~\cite{chuan05:S2C_tune}:
 \ba
 \label{eq:shift_M}
 {\mathcal A}_{xy} &=&\delta_{xy}\left[1/(2\kappa_{max})
 +\rho_t \sum^3_{i=1} \sigma_{0i} {\mathcal F}_{0i}
 +\rho_s (\sigma_{12}{\mathcal F}_{12} +\sigma_{23}{\mathcal F}_{23}
 +\sigma_{31}{\mathcal F}_{31})\right]
 \nonumber \\
 &-&\sum_{\mu} \eta_{\mu} \left[
 (1-\gamma_\mu) U_\mu(x) \delta_{x+\mu,y}
 +(1+\gamma_\mu) U^\dagger_\mu(x-\mu) \delta_{x-\mu,y}\right]
 \;\;.
 \ea
 The coefficients in the matrix are given explicitly by:
  \ba
 \label{eq:parameters_TI}
 \eta_i &=&{\nu \over 2u_s} \;\;, \;\;
 \eta_0={\xi\over 2} \;\;,
 \;\;\sigma={1 \over 2\kappa}-{1\over 2\kappa_{max}}\;\;,
 \nonumber \\
 \rho_t &=& \nu{(1+\xi)\over 4u^2_s} \;\;, \;\;
 \rho_s = {\nu \over 2u^4_s} \;\;.
 \ea
 In this letter, we present our preliminary quenched results for
 the meson spectra using the above improved quark action on anisotropic
 lattices.

 \section{Simulation Results}

 The whole procedure of the calculation involves three major steps:
 In the first step, we determine the optimal value of
 the bare speed of light parameter $\nu$, using energy-momentum dispersion
 relations of pseudo-scalar mesons; In the second step, the masses of the
 mesons that we are interested in are extrapolated/interpolated to
 the optimal value of $\nu$; In the third step, we perform chiral and
 continuum extrapolations as usual.

 The lattices in this study are of size
 $6\times9\times12\times50$ for $\beta=2.4, 2.6$ and
 $8\times 12\times 16\times 50$ for $\beta=2.8$. The spatial
 lattice spacing $a_s$ roughly range from $0.13fm$ and $0.23fm$ in physical
 units. The aspect ratio is $\xi=a_s/a_t\simeq\xi_0=5$ for all lattices.
 For gauge field configurations at
 a given value of $\beta$, $4$ different values of
 the bare speed of light parameter $\nu$ and $12$
 different values of the hopping parameter $\kappa$
 are studied. The range of $\kappa$ roughly covers
 the quark mass from the strange to the charm.
 About $300$ configurations are used for each parameter set in this study.

 The energy values of meson with definite three-momentum $\bk$
 (including zero-momentum) is obtained from their respective correlation
 functions by finding the plateaus in their effective mass plots.
 The errors for the data points in the effective mass plots are analyzed
 using the standard jack-knife method.
 We obtain from the pseudo-scalar and vector channel the corresponding energy
 $E_{PS/V}(m_1,m_2,\nu_1,\nu_2,\bk)$, where $m_1$ and $m_2$
 are the bare quark mass parameters defined by:
 $ m\equiv 1/(2\kappa)-1/[2\kappa_{cr}(\nu)]$.
 The value of $1/[2\kappa_{cr}(\nu)]$ is obtained by
 the condition of vanishing pseudo-scalar meson mass
 for each $\nu$. Although for different values of $\nu$,
 the set of values for $m$ turns out to be slightly
 different, we always interpolate/extrapolate them to
 the same set of values~\cite{chuan05:S2C_tune}.

 To obtain the optimal value of $\nu$,
 the values of $E^2_{PS}(m,m,\nu,\nu,\bk)$ are fitted
 linearly against the three-momentum squared $\bk^2$ in the
 low-momentum region:
 \be
 \label{eq:dis}
 E^2_{PS}(m,m,\nu,\nu,\bk) = M^2_{PS}(m,m,\nu,\nu)
 + Z_{PS}(m,m,\nu,\nu)\bk^2\;\;.
 \ee
 The condition: $Z_{PS}[m,m,\nu_{opt}(m),\nu_{opt}(m)]=1$ then
 yields the optimal value of $\nu$ as a function
 of the bare quark mass parameter: $\nu_{opt}(m)$.
 In practice, we use the values of $Z_{PS}(m,m,\nu,\nu)$ at
 different $\nu$ to perform a linear extrapolation/interpolation
 in $\nu$ for each value of $m$.

 We now proceed to the second step of our study, namely to
 extrapolate/interpolate the meson mass values to the optimal
 values of $\nu$ obtained from the pseudo-scalar dispersion
 relations. This is achieved by a linear
 extrapolation/interpolation in $\nu$, or equivalently
 in the quantity $Z_{PS}(m,m,\nu,\nu)$.

 Similar procedure discussed above can be repeated for
 mesons consisting of a quark and an anti-quark with different mass.
 We interpolate/extrapolate the meson masses to the optimal bare speed
 of light parameter. The outcome of this procedure are the quantities
 of $M_{PS/V}^2(m_1,m_2,\nu_{opt}(m_1),\nu_{opt}(m_2))$. Since these
 quantities only depend on the bare quark mass parameter pairs $(m_1,m_2)$,
 we denote them by $\bar{M}_{PS/V}^2(m_1,m_2)$.

 In the third step of the analysis, we perform the chiral
 extrapolations. First, we determine the physical strange
 and charm bare quark mass parameters: $m^{(phy)}_s$ and $m^{(phy)}_c$.
 In this work, we use the mass of the vector meson with
 the same flavor of quarks to determine $m^{(phy)}_s$ and $m^{(phy)}_c$.
 The results of $M_V^2(m,m,\nu_{opt}(m),\nu_{opt}(m))$
 are fitted quadratically versus the quark mass parameter:
 \be
 \label{eq:chiral_pion}
 a^2_tM^2_{V}[m,m,\nu_{opt}(m),\nu_{opt}(m)]= A+Bm+Cm^2\;.
 \ee
 We perform two such quadratic fits, one in the low quark mass
 region, the other in heavy quark mass region.
 The scale of the lattice is set by $r_0=0.5$fm (the
 so-called Sommer scale).
 Then the value of $m^{(phy)}_s$ and $m^{(phy)}_c$ are obtained by
 requiring the vector meson mass obtained from the simulation to
 exactly reproduce the mass of the physical
 $\Phi$ and  the physical $J/\psi$ meson mass.
 \begin{figure}[tb]
 \begin{center}
 \includegraphics[height=12.0cm,angle=0]{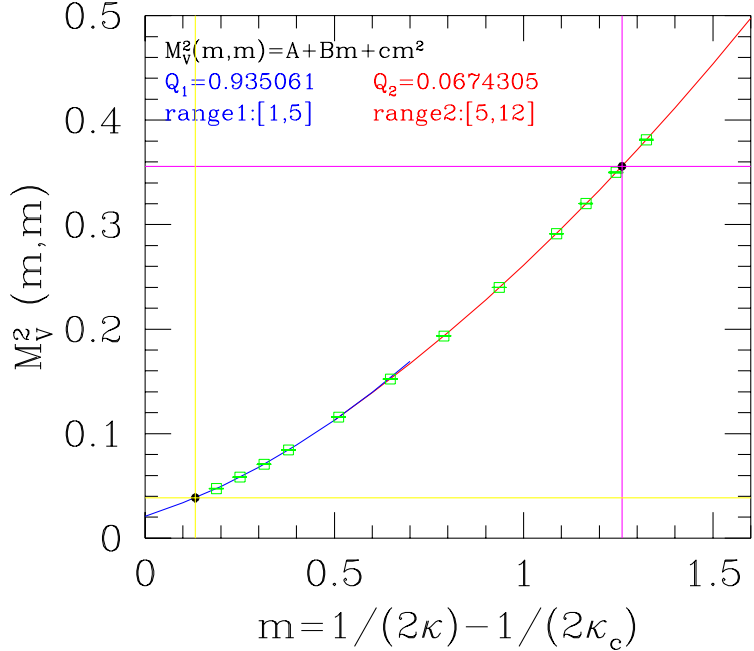}
 \end{center}
 \caption{Quadratic extrapolation/interpolation of
 $M^2_V(m,m,\nu_{opt}(m),\nu_{opt}(m))$ versus
 the bare quark mass $m$ at $\beta=2.6$. The two curves.
 are the fits in the light and heavy
 quark mass regions, respectively. The
 corresponding fitting qualities and ranges are also
 shown in the figure. The fitting ranges are self-adjusted
 to yield minimum $\chi^2$ per
 degree of freedom. The horizontal and
 vertical lines indicate the value for the physical $\Phi$ and
 $J/\psi$ meson and the fixed value of bare quark mass
 $m_c^{(phy)}$ and $m_c^{(phy)}$, respectively.}
 \label{fig:xcxs.plot}
 \end{figure}
 Fig.~\ref{fig:xcxs.plot} shows the quadratic extrapolations of
 $M^2_V(m,m,\nu_{opt}(m),\nu_{opt}(m))$ versus the bare quark mass
 $m$ at $\beta=2.6$. The physical $\Phi$  and $J/psi$ mass are also lined out
 by horizontal lines in the figure. The
 intersection point of the lower/higher horizontal line
 representing $a^2_tM^2_{\Phi}$/$a^2_tM^2_{J/\psi}$ and
 the quadratic fitting curve yields the estimate for the physical
 strange quark mass parameter $m^{(phy)}_s$ and $m^{(phy)}_c$, respectively.

 Having determined the values of $m^{(phy)}_s$ and $m^{(phy)}_c$,
 we then make quadratic extrapolations/interpolations for all the meson
 mass squared ($\bar{M}_{PS/V}^2(m_1,m_2)$) versus the quark mass parameters
 and extrapolate/interpolate them to their physical values
 $m^{(phy)}_{s/c}$, depending on which type of meson is being
 considered. The final results are the mass values of
 various charmed mesons and charmonium states at different $\beta$.
 These results are listed in Table~{tab:spectra}.
 \footnote{The mass of $J/\psi$ (not included in Table~\ref{tab:spectra})
 is by definition that of the experimental value
 since we have used the mass of $J/\psi$ to determine the bare charm quark mass. }
 \begin{table}[htb]
 \caption{\label{tab:spectra} Extracted physical mass (in unit of MeV)
 of vector and pseudo scalar meson made up of one charm quark and another
 quark, for different $\beta$ and continuum limit. For comparison, we also
 list the experimental values for the mesons in the last column.}
 \begin{center}
 \begin{tabular}{|c|c|c|c|c|c|}
 \hline
 $\beta$                  & $2.8$      & $2.6$      & $2.4$      & continuum limit & Experiment\\
 \hline\hline
 $m(D)$                   & $1835(35)$ & $1862(5)$ & $1860(9)$   & $1847(68)$ & $1869$\\
 \hline
 $m(D_s)$                 & $1976(21)$ & $1993(3)$ & $1959(5)$   & $2164(39)$ & $1969$\\
 \hline
 $m(\eta_c)$              & $2992(2)$  & $3037(1)$ & $3020(10)$  & $2843(23)$  & $2980$\\
 \hline
 $m(D^\ast)$              & $2015(27)$ & $1965(6)$ & $1993(33)$  & $2046(131)$ & $2007$\\
 \hline
 $m(D_s^\ast)$            & $2140(18)$ & $2087(3)$ & $2080(13)$  & $2192(67)$ & $2112$\\
 \hline
 \end{tabular}
 \end{center}
 \end{table}

 \begin{figure}[tb]
 \begin{center}
 \includegraphics[height=12.0cm,angle=0]{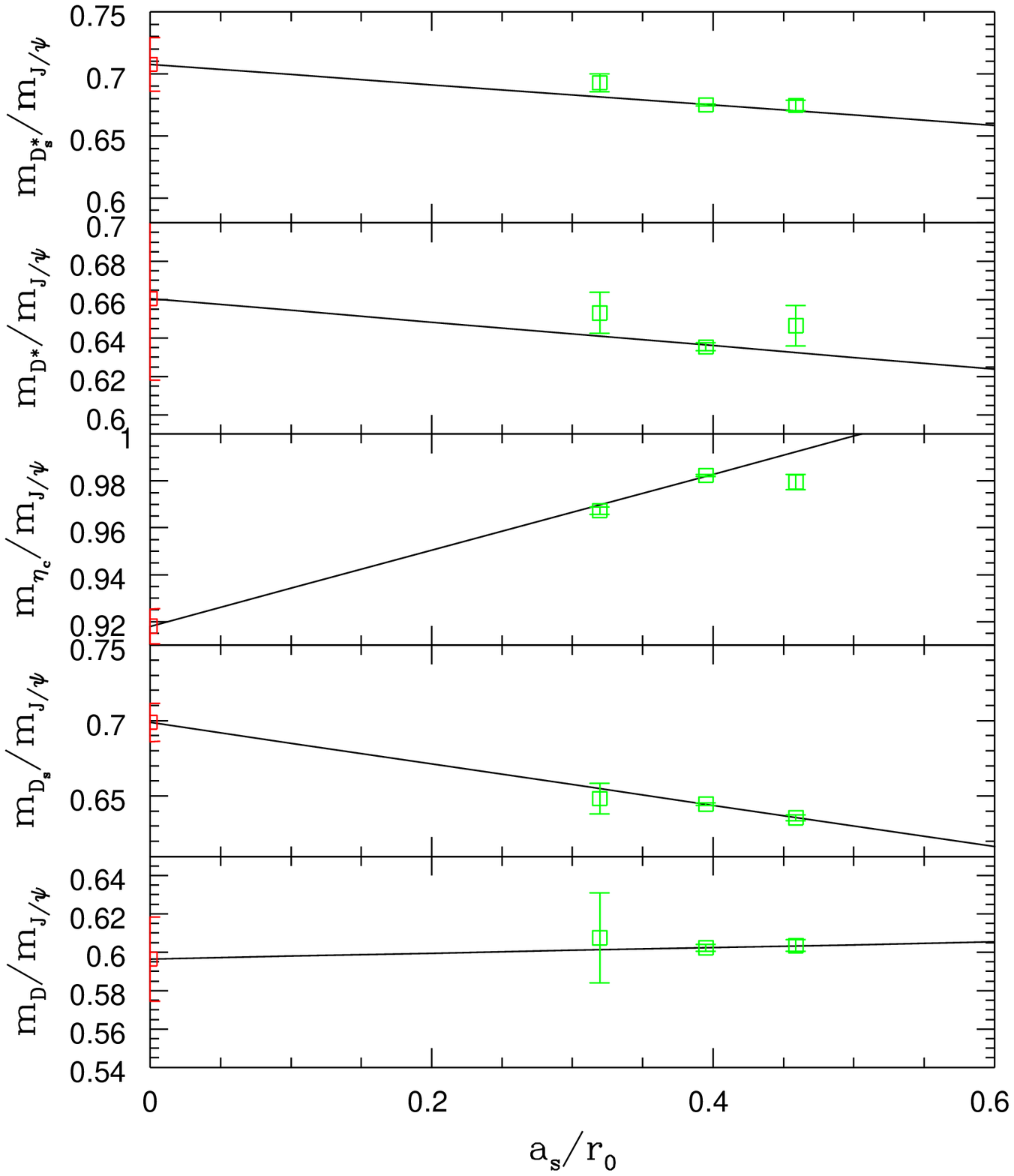}
 \end{center}
 \caption{ Continuum limit extrapolation for each meson
 mass over $J/\psi$ meson mass. Each panel indicates
 different meson. Data points (shown by green points)
 in the plot are ratios of the mass of each meson to
 the mass of $J/\psi$ for different $\beta$ (corresponding to different
 value of $a_s/r_0$). The red points represent the physical mass ratios
 at the continuum limit.}
 \label{fig:mratio_beta.plot}
 \end{figure}
 Finally, the results of the meson masses in the continuum are obtained by
 extrapolate their values at finite lattice spacings to the continuum limit.
 In Fig.~\ref{fig:mratio_beta.plot}, we show the mass ratios of various
 mesons over the mass of the $J/\psi$ versus lattice spacing $a_s$.
 Linear fits are performed for these mass ratios and the results are indicated
 by the straight lines. The extrapolated results are also indicated
 by the points at $a_s=0$.
 The numerical results are also listed in Table~\ref{tab:spectra}.
 The fitting quality for most cases are
 reasonable. However, since we have only $3$ different values of
 $\beta$ for this preliminary study, the continuum limit extrapolations
 suffer big errors. Needless to say, more points at different
 values of $a_s$, preferably at smaller $a_s$, will help to
 improve the final result at the continuum limit.

 To compare with the experiment, we have also tabulated the
 results from Particle Data Group (PDG) in the last
 column of Table~\ref{tab:spectra}. It is seen that the
 qualitative agreement is good, especially for
 $D$, $D^\ast$ and $D^\ast_s$ mesons. The mass values of the
 $D_s$ and $\eta_c$ are away from the experimental value
 by $5\sigma$. These disagreements might
 come from the fact that our lattice spacing is
 still somewhat large and the continuum extrapolation
 is not very well controlled. Of course, quenching effects are
 also present which might be important for the fine
 structure of the spectrum.

\section{conclusion}

 In this paper, we present a preliminary study
 of the quenched meson spectra for the charmed mesons and
 charmonium states. To perform such a calculation using
 the tadpole improved anisotropic Wilson quark action, we
 tune the bare speed of light parameter $\nu$ in a quark mass dependent
 way with quark mass values ranging from the strange to the charm.
 The optimal values of $\nu$ are obtained for various
 values of $\beta$ using the pseudo-scalar meson
 dispersion relations.
 Using the tadpole improved anisotropic Wilson action with
 these optimized parameters, we determine the ground state
 meson spectra for $D$, $D_s$, $\eta_c$, $D^{\ast}$, $D_s^{\ast}$ and
 $J/\psi$ mesons. The results are extrapolated towards
 the continuum limit where qualitative agreement with
 the experiment is found.
 With the method outlined in this preliminary study,
 we plan to carry out spectrum calculations of other
 excited meson states, baryon states and possibly
 hybrid states involving a charm quark in the future.


\end{document}